# Measurement of Spin Polarization by Andreev Reflection in Ferromagnetic In$_{1-x}$Mn$_x$Sb Epilayers

R. P. Panguluri, and B. Nadgorny[a]

Department of Physics and Astronomy, Wayne State University, Detroit, MI 48201

T. Wojtowicz,

Department of Physics, University of Notre Dame, Notre Dame, IN 46556 and

Institute of Physics, Polish Academy of Sciences, 02-668 Warsaw, Poland

W.L. Lim, X. Liu, and J.K. Furdyna,

Department of Physics, University of Notre Dame, Notre Dame, IN 46556

Abstract:

We carried out Point Contact Andreev Reflection (PCAR) spin spectroscopy measurements on epitaxially-grown ferromagnetic In$_{1-x}$Mn$_x$Sb epilayers with a Curie temperature of ~9K. The spin sensitivity of PCAR in this material was demonstrated by parallel control studies on its non-magnetic analog, In$_{1-y}$Be$_y$Sb. We found the conductance curves of the Sn point contacts with In$_{1-y}$Be$_y$Sb to be fairly conventional, with the possible presence of proximity-induced superconductivity effects at the lowest temperatures. The experimental *Z*-values of interfacial scattering agreed well with the estimates based on the Fermi velocity mismatch between the semiconductor and the superconductor. These measurements provided control data for subsequent PCAR measurements on ferromagnetic In$_{1-x}$Mn$_x$Sb, which indicated spin polarization in In$_{1-x}$Mn$_x$Sb to be 52 ± 3%.

[a] Electronic mail: nadgorny@physics.wayne.edu

The recent emergence of III–Mn–V dilute ferromagnetic semiconductor alloys, such as $In_{1-x}Mn_xAs$[1] and $Ga_{1-x}Mn_xAs$[2], has already led to a number of exciting results relevant to spintronics applications. The ability to fabricate tunneling magneto-resistance (TMR) devices[3] naturally integrated with the technologically important semiconductors, such as GaAs, makes these materials especially attractive. The need to increase the operating temperatures of spintronics devices to room temperature has stimulated extensive studies of $Ga_{1-x}Mn_xAs$ epilayers, where the highest Curie temperature is now close to 190K[4]. The studies of $Ga_{1-x}Mn_xN$ [5] and other dilute magnetic semiconductors with smaller lattice constants and larger effective valence-band masses have resulted in even higher Curie temperatures. While the ferromagnetic $In_{1-x}Mn_xSb$ alloy[6,7,8] is much less explored, it has – in spite of its lower Curie temperature – significant potential for applications in infrared *spin-photonics* and in spin-dependent transport devices due its lighter holes, small energy gap, and much higher carrier mobility than other III-Mn-V ferromagnetics.

The efficiency of most of spintronic devices, such as giant magneto-resistance (GMR) or TMR junctions, depends on the *transport* spin polarization of carriers in ferromagnetic materials. The point contact Andreev reflection (PCAR)[9,10] has recently been introduced as an effective technique for measuring the transport spin polarization $P_c$ of metals and metallic oxides[11], $P_c = \frac{<N_\uparrow(E_f)V_{f\uparrow}> - <N_\downarrow(E_f)V_{f\downarrow}>}{<N_\uparrow(E_f)V_{f\uparrow}> + <N_\downarrow(E_f)V_{f\downarrow}>}$, in the ballistic limit. PCAR is based on the difference in the Andreev reflection process in normal metal/superconductor (N/S)[12] and in ferromagnet/superconductor (F/S) contacts[13]. While at the N/S interface all quasi-particles with the energies $eV \leq \Delta$ (where $\Delta$ is the superconductor energy gap) incident from a non-magnetic material onto a superconductor are converted into Cooper pairs, at the F/S interface uncompensated quasi-particles are unable to propagate, thus reducing the conductance. This strongly affects the overall character of the conductance curves, d*I*/d*V*, which can then be related to the degree of the spin polarization of the ferromagnet[14].

The key to extending the PCAR technique to FSm/S interface is the requirement of the *high junction transparency*, which is often limited by the native Schottky barriers present at most semiconductor/superconductor (Sm/S) interfaces. Most of the experimental work on Andreev reflection in semiconductors, starting with the pioneering work of Kastalsky *et al.*[15], has been done in a 2D configuration. While the Andreev process for metallic F/S interfaces in spite of some remaining theoretical questions[16] is fairly well established, Sm/S interface is relatively unexplored[17], and a number of important theoretical questions still remain unanswered. In particular, the complete theory of Andreev reflection in magnetic semiconductors – which should include the effects of spin-orbit interaction, disorder, and possibly also the effects of exchange interaction – is yet to be developed.

In our recent work, we have demonstrated a conventional 3D Andreev reflection in a non-magnetic semiconductor using low temperature (LT)-GaAs doped up to a very high level with Be, that produced the free hole concentration as high as p = $8 \times 10^{20}$ cm$^{-3}$. From the estimates of the junction transparency we have concluded that the effects of the Schottky barrier are negligible, which in this case is not surprising, since the heavy doping dramatically reduces the effective barrier thickness. These measurements suggested, therefore, that the PCAR spectroscopy could be successfully applied to other dilute magnetic semiconductors, at least to those with very high carrier concentrations. However, the first Andreev reflection studies on FSm/S junctions using $Ga_{1-x}Mn_xAs$ epilayers with carrier concentrations comparable to that of our $Ga_{1-y}Be_yAs$, carried out both by us[18] and by the Florida State group[19], indicated unconventional behavior, possibly arising from some not yet understood interaction between the superconductor and the magnetic semiconductor. These results suggest that the interpretation of the PCAR experiments on ferromagnetic semiconductors would be facilitated by direct comparison with Andreev reflection measurements carried out on analogous non-magnetic semiconductors. Additionally, to separate different mechanisms contributing to the observed effects, it is important to extend Anreev reflection studies to other members of the III-Mn-V family of ferromagnetic semiconductors, especially to those with very different physical parameters compared to $Ga_{1-x}Mn_xAs$. In this respect $In_{1-x}Mn_xSb$ – with the effective masses lowest among the III-Vs, and hence the highest mobilities – is an excellent candidate, as it is expected to form high-transparency FSm/S junctions. Moreover, it is fortunate that its nonmagnetic analog, $In_{1-y}Be_ySb$, needed for comparative studies can also be grown by the low temperature molecular beam epitaxy.

In this letter we report PCAR spin polarization measurements of $In_{1-x}Mn_xSb$ epilayers with the Curie temperature ~9K. In order to facilitate the interpretation of the $In_{1-x}Mn_xSb$ results, these measurements are accompanied by detailed comparison with the Andreev reflection data obtained on its non-magnetic analog, $In_{1-y}Be_ySb$.

The growth of both $In_{1-x}Mn_xSb$ and $In_{1-y}Be_ySb$ films was carried out by low-temperature (LT) molecular beam epitaxy, as described in detail in Refs. 6 and 7. The In and Mn (Be) fluxes were supplied from standard effusion cells, while $Sb_2$ flux was produced by an Sb cracker cell. Hybrid (100) CdTe/GaAs and InSb/AlSb/GaAs substrates were used for the growth. Prior to depositing each $In_{1-x}Mn_xSb$ or $In_{1-y}Be_ySb$ film we grew a 100 nm LT- InSb buffer layer at 210 °C which, as seen from a well-resolved RHEED pattern, provided a flat surface for subsequent deposition. The substrate was then cooled to 170 °C for the growth of a 230-nm-thick LT- $In_{1-x}Mn_xSb$ or LT-$In_{1-y}Be_ySb$ epilayer. A 3:1 $Sb_2$:In beam equivalent pressure ratio was used for the growth of both InSb-based systems.

In this paper we studied several $In_{1-x}Mn_xSb$ epilayer with $x \approx 0.03$. The specimens had a Curie temperature of approximately 9K, as determined from the temperature-dependence of the remanent magnetization (see Fig.1a), and from the series of hysteresis curves taken at different temperatures (Fig. 1b). For the *non-magnetic* analog to be used as the control sample we chose an epilayer of $In_{1-y}Be_ySb$ with $y = 0.05$, selected in such a way that its free hole concentration $p = 1.5 \times 10^{20}$ cm$^{-3}$ was close to that of the ferromagnetic film (p ~$2 \times 10^{20}$ cm$^{-3}$, see Ref. 6).

Mechanically sharpened Sn superconductor tips were used for all the point contact measurements reported in this study. A contact was established between the sample and the Sn tip when the probe was completely immersed into a liquid He bath. The conductance was measured by the standard four-terminal technique using lock-in detection at 2kHz. Conductance curves were analyzed by means of the modified BTK model[14] with two fitting parameters: the spin polarization $P_c$ and the dimensionless interfacial scattering parameter $Z$. Details of the measurement and fitting procedures are given in Refs.20 and 14, respectively.

In contrast to the behavior observed in $Ga_{1-x}Mn_xAs$ results[18] all measurements of Sn/In$_{1-x}$Mn$_x$Sb point contacts indicate a fairly conventional Andreev reflection behavior of a ferromagnet/superconductor junction. Representative d$I$/d$V$ curves for two different contacts at 1.2K and 1.6K are shown in Fig. 2. The superconductor gap value used for the analysis of all our data, $\Delta(0) \sim 0.52$mV, was close to the bulk value for Sn. The gap at higher temperatures was obtained from the BCS $\Delta(T)$ dependence, with the critical temperature of Sn also quite close to its bulk value, $T_c \sim 3.7$K. All of the experimental d$I$/d$V$ curves display a characteristic dip at zero bias, which is consistent with the suppression of Andreev reflection due to the spin polarization of In$_{1-x}$Mn$_x$Sb.

The value of the spin polarization $P_c$ for In$_{1-x}$Mn$_x$Sb was obtained after analyzing each of the d$I$/d$V$ curves for every contact at the corresponding temperatures by using the theory of Ref. 14. The average spin polarization was found to be $P_c \sim 52 \pm 3\%$. The typical values of $Z \leq 0.25$ indicate fairly high junction transparencies. It is important to note that even with nominally no-barrier contact some quasiparticle reflection is always present due to the so-called Fermi velocity mismatch $r = v_S/v_{Sm}$ between a superconductor and a semiconductor, where $v_S$ and $v_{Sm}$ are the Fermi velocities in the superconductor and the semiconductor, respectively. To estimate this Fermi velocity mismatch in both the Sn/In$_{1-x}$Mn$_x$Sb and Sn/In$_{1-x}$Be$_x$Sb contacts, we used the value of the density-of-states effective mass in InSb[21] $m^* = 0.25\ m_o$ and the free hole concentration p = 2x10$^{20}$ cm$^{-3}$ to estimate the Fermi velocities in InSb. Taking the value of the Fermi velocity of Sn to be 1.9x10$^8$ cm/s, we obtain $r = 2.3$, which yields a minimum $Z$-value of $\sim 0.4$. These estimates are in fairly good agreement with the experimental $Z$-values obtained for In$_{1-x}$Mn$_x$Sb and for Sn/In$_{1-x}$Be$_x$Sb described below.

From the known values of the hole density and the resistivity $\rho \sim 0.2$ m$\Omega$-cm, we can also estimate the low temperature mean free path $L$ for light and heavy holes to be ~60 nm and ~15nm respectively. The size of the contact $d$ can then be estimated from the Sharvin formula $R_n = (4\rho L/3\pi d^2 + \rho/2d)(1+Z^2)$, where $R_n$ is the contact resistance. For the typical values of the contact resistance $R_n \sim 60\Omega$, we have obtained the contact size $d \sim 25$nm, indicating that our measurements were done essentially in the ballistic regime, $L \geq d$.

To test the reliability of our measurements on the spin-polarized $In_{1-x}Mn_xSb$ system, we have studied a large number of non-magnetic $Sn/In_{1-x}Be_xSb$ junctions. A series of characteristic conductance curves for one of these contacts at different temperatures is shown in Fig. 3. In contrast to the $Sn/In_{1-x}Mn_xSb$ contacts, these curves exhibit a higher conductance at zero bias compared to their normal conductance above the gap. This behavior, together with the increase of the amplitude of the zero-bias conductance at lower temperatures, is consistent with Andreev reflection for high transparency junctions. The inset in the figure shows the fitted conductance curves at different temperatures. The $Z$-values obtained for all fitted curves have a *temperature-independent* value of approximately 0.39 ± 0.05, in good agreement with the $Z$-values obtained for the analogous InMnSb system, as well as with the estimates based on the Fermi velocity mismatch obtained above. At lower temperature, however, we consistently observe some discrepancy between the experimental curves and the best fit, as can be seen from the upper curve in Fig. 3. One of the possible explanations could be the presence of proximity-induced superconductivity in the $In_{1-x}Be_xSb$ film. This effect, which was especially pronounced in some of the lowest-resistance contacts, can be found in the literature and is often interpreted either as a second order proximity-induced

Josephson effect[22], or as a first order Josephson effect associated with phase-slip center[23].

In summary, we have measured the transport spin polarization of the dilute ferromagnetic semiconductor $In_{1-x}Mn_xSb$. Simultaneously we have made a detailed study of the analogous system $In_{1-y}Be_ySb$, which served as a non-magnetic control material for the PCAR measurements on InMnsb. The measurements in both magnetic and non-magnetic systems demonstrate fairly conventional Andreev reflection in high-transparency junctions, with no measurable DOS broadening and with interface transparencies, $Z$ close to minimum values estimates from Fermi velocity mismatch. The spin polarization of $In_{1-x}Mn_xSb$ was determined to be 52 ± 3%. The measurements on $In_{1-y}Be_ySb$ indicate a possible influence of the proximity effect on the junction properties. While the behavior of the Sn/$In_{1-y}Be_ySb$ contacts is similar to the behavior of the Sn/$Ga_{1-y}Be_yAs$ system[18], there is a marked difference between the behavior of Sn contacts with $Ga_{1-x}Mn_xAs$ and with $In_{1-x}Mn_xSb$. Narrower band gap as well as a much higher carrier mobility, characteristic of $In_{1-x}Mn_xSb$ as compared to $Ga_{1-x}Mn_xAs$, may be just some of a number of important parameters affecting the physics of Andreev reflection in these systems, and it is our hope that this experimental study will stimulate the development of the theory of Andreev reflection for III-Mn-V ferromagnetic semiconductors.

We thank I.I. Mazin, A. Petukhov, and I. Zutic, for useful discussions. This work at Wayne State supported by DARPA through ONR grant N00014-02-1-0886 and NSF Career grant (B.N.). The work at Notre Dame supported by the DARPA SpinS Program.

Figure Captions.

Fig. 1. Field dependence of the magnetization in the $In_{1-x}Mn_xSb$ epitaxial film with $x = 0.028$. The data were collected for a series of temperatures with the applied field perpendicular to the layer plane. The inset shows the temperature dependence of the remanent magnetization, indicating the Curie temperature of ~ 9K.

Fig. 2. Typical normalized conductance curves for two different Sn superconductor contacts with $In_{1-x}Mn_xSb$ epitaxial films: (a) Contact resistance $R_c = 57\Omega$, $T = 1.2K$; $\Delta = 0.52$mV; fitting parameters: $Z = 0.19$ and $P = 54\%$; (b) Contact resistance $R_c = 36\Omega$, $T = 1.6K$; $\Delta (1.6K) = 0.5$ mV; fitting parameters: $Z = 0.20$ and $P = 52\%$.

Fig. 3. Normalized conductance as a function of voltage for a Sn superconductor contact with an $In_{1-y}Be_ySb$ epitaxial film (contact resistance of $R_c = 63\Omega$) for a series of temperatures. The inset shows fitting of the experimental data (solid curve) for three different temperatures. The $Z$-parameter for all the fitted curves is $0.39 \pm 0.05$.

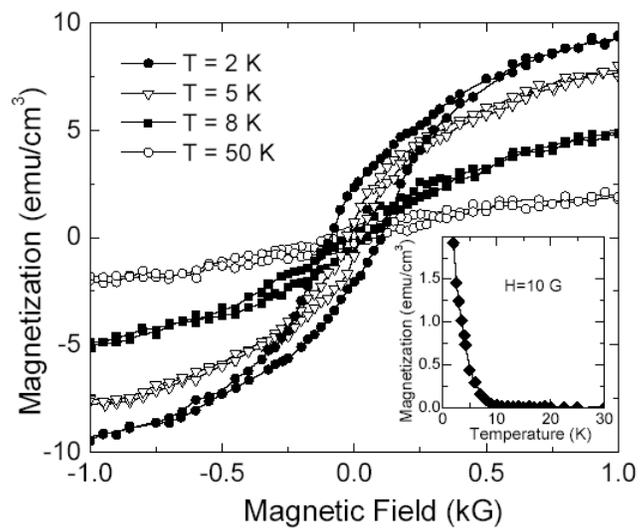

**Figure 1, Panguluri et al.**

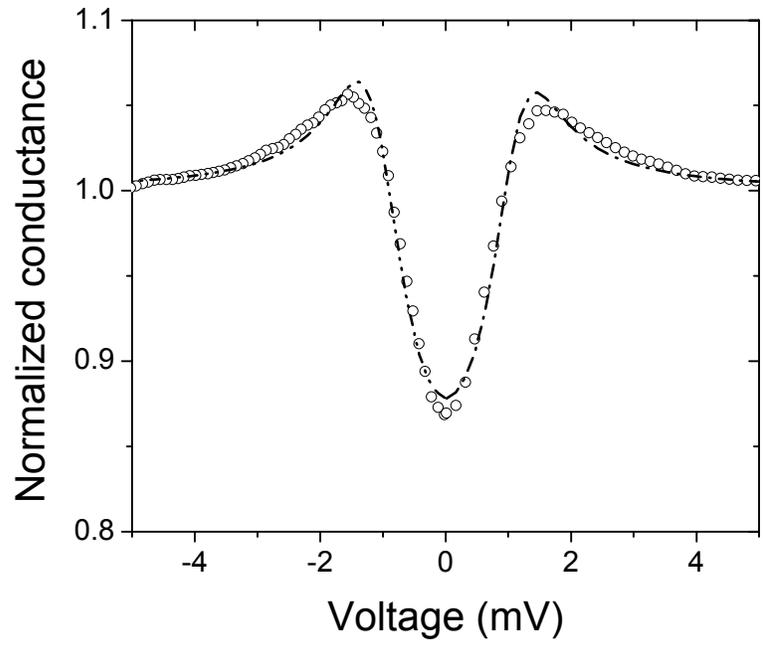

(a)

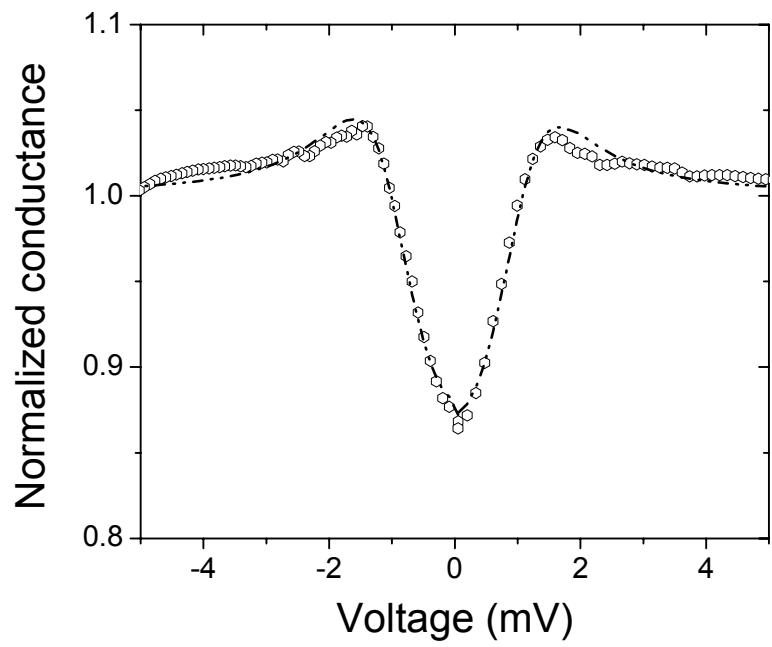

(b)

**Figure 2 (a,b), Panguluri et al.**

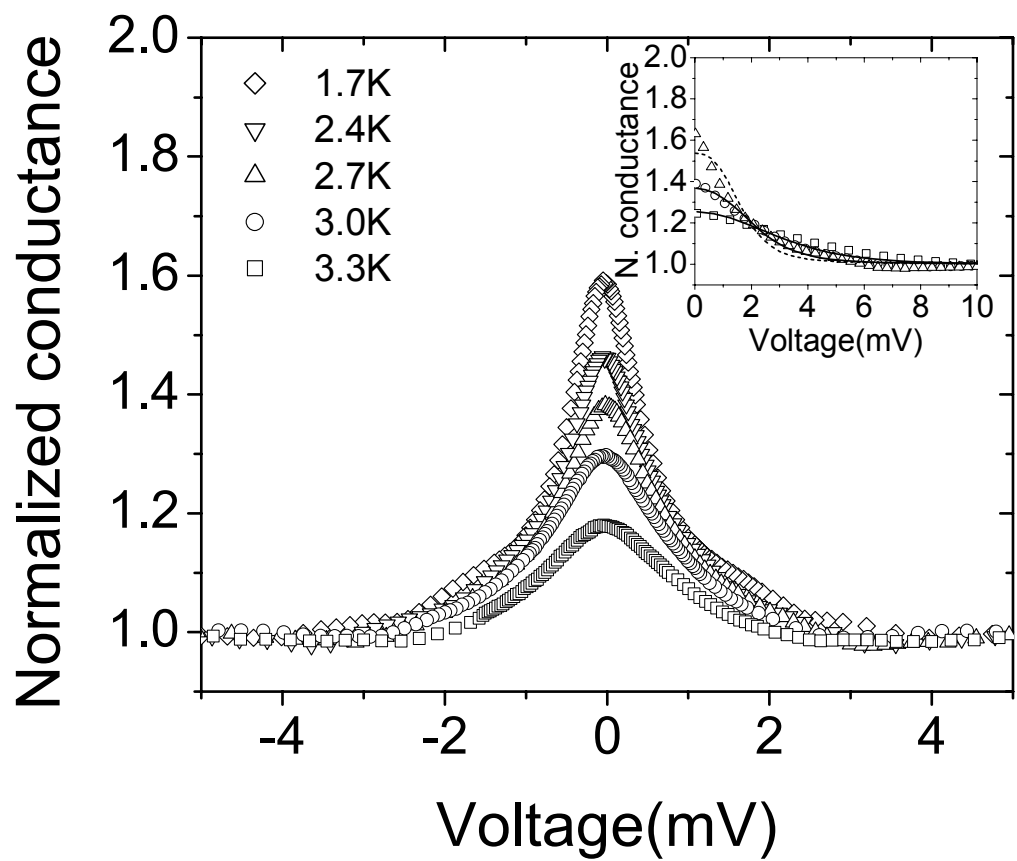

**Figure 3, Panguluri et al.**